\documentclass{article}
\begin{document}
\vbox{\vspace{6mm}}
\renewcommand{\thefootnote}{\fnsymbol{footnote}}
\setcounter{footnote}{1}
\begin{center} 
{\Large  \bf Coherent States from Combinatorial Sequences}
\footnote{Talk presented at the Second Conference on Quantum Theory and Symmetry, Cracow, Poland, 17--21 July 2001} 
\\[7mm]
K.A Penson\footnote{e-mail : penson@lptl.jussieu.fr} and 
A.I Solomon\footnote{Permanent address: Quantum Processes Group, Open University, Milton Keynes, MK7 6AA, United Kingdom.\\e-mail : a.i.solomon@open.ac.uk} 
\\[3mm]
Universit\'{e} Pierre et Marie Curie, Laboratoire de Physique Th\'{e}orique des Liquides,  75252 Paris Cedex 05, France.
\end{center}

\begin{abstract}
We construct coherent states using sequences of combinatorial numbers such as various binomial and trinomial numbers, and Bell and Catalan numbers.  We show that these states satisfy the condition of the {\em resolution of unity} in a natural way.  In each case the positive weight functions are given as solutions of associated Stieltjes or Hausdorff moment problems, where the moments are the combinatorial numbers. 
\end{abstract}

We describe the construction of coherent states which are generalizations of the standard coherent states defined for complex $z$ by\cite{kla}
\begin{equation}
|z\rangle_0 = e^{-\frac{|z|^2}{2}}\sum_{n=0}^\infty \frac{z^n}{\sqrt{n!}}|n\rangle  ,
\label{cs}
\end{equation}
where $|n\rangle$ are eigenfunctions of a Hermitian operator $H$
 (usually the system Hamiltonian) with
\begin{equation}
H|n\rangle =\varepsilon_n|n\rangle \; \; \; \; \langle n|n' \rangle = \delta_{n,n'},\; \; \; \; n=0,1,\ldots.
\end{equation}
The generalization consists in defining new states 
\begin{equation}
|z\rangle_c = {\cal N}{_c}^{-\frac{1}{2}}(|z|^2) \sum_{n=0}^\infty {\frac{z^n}{\sqrt{c(n)}}}|n\rangle ,
\label{ccs}
\end{equation}
where
\begin{equation}
{\cal N}{_c}(x) = \sum_{n=0}^\infty {\frac{x^n}{c(n)}}, \;  \; x\leq R.
\label{norm} \end{equation}
In Eq.(\ref{ccs}) and Eq.(\ref{norm}), ${\cal N}{_c}(x) \; (x\equiv|z|^2)$ is the normalization of $|z\rangle$, and $R\leq \infty$ is its radius of convergence. We shall assume that in Eq.(\ref{ccs}) the positive numbers $c(n)$ for $n=0,1, \ldots, $ arise from sequences of combinatorial numbers; that is, integers which count the characteristic properties of an ensemble of $n$ objects ($c(0)=1$ by convention). In this note we shall refrain from providing the precise combinatorial interpretation of the sequences used, but shall rather limit ourselves to giving their formulae. For more details of the $c(n)$ we refer the interested reader to the classical literature \cite{com,sta,slo}.

In general, the states $|z\rangle_c$ are not orthogonal
 ($\langle z|z' \rangle_c \neq 0$); the resolution of unity condition is given by the following sum of weighted non-orthogonal projectors
\begin{equation}
\int\!\!\int d^2 z |z\rangle_c W_c(|z|^2)\;_{c}\langle z| = I = \sum_{n=0}^{\infty} |n\rangle \langle n|,
\label{res}
\end{equation}
where in Eq.(\ref{res}) the integration is restricted to the part of the complex plane where the normalization converges. Equations (\ref{norm}) and (\ref{res}) may be succinctly written as
\begin{equation}
\int_0^R x^n {\tilde{W}}_c(x) dx = c(n), \; \; \; \; \; \; n=0,1,\ldots,
\label{res1} \end{equation}
with ${\tilde{W}}_c(x) = \pi W_c(x)/ {\cal N}{_c}(x) >0.$
For $R=\infty$, Eq.(\ref{res1}) is the Stieltjes, and for $R< \infty $, the Hausdorff moment problem, with moments $c(n)$. 

The motivation for choosing the combinatorial sequences $\{c(n)\}$ as input moments to Eq.(\ref{ccs}) is that for many such sequences solutions can be explicitly constructed by considering Eq.(\ref{res1}) as a Mellin transform; see \cite{six,six1,kla1,que} for related instances.  
For example, in those cases where the $c(n)$ may be expressed solely in terms of gamma functions, the use of properties of Meijer's G function is instrumental in obtaining solutions of Eq.(\ref{res1}). In what follows we shall list a number of sequences $\{c(n)\}$ and simply quote the solutions of Eq.(\ref{res1}), specifying $R$ in each case: 
\begin{enumerate}
\item $c(n) =(2n)! \, \, ; \, \, \, \, {\tilde{W}}_1(x)  = \frac{1}{2} e^{-\sqrt x}/\sqrt x ;\,\,\, R=\infty$.
\item $c(n) =(2n)!/n! \, \,; \, \, \, \, {\tilde{W}}_2(x)  = \frac{1}{2\sqrt{\pi}} e^{-\frac{x}{4}}/\sqrt{x} ;\,\,\, R=\infty$.
\item $c(n) =\left(\begin{array}{c}2n\\n\end{array}\right)\, \,$ (so-called central binomial coefficients) ;$\\ {\tilde{W}}_3(x)  = \frac{1}{\pi} [x(4-x)]^{-\frac{1}{2}} ;\,\,\, R=4$.
\item $c(n) =\left(\begin{array}{c}2n\\n\end{array}\right)/(n+1)$ (so-called Catalan numbers) ;$\\ {\tilde{W}}_4(x)  = \frac{1}{\pi} [(4-x)/x]^{\frac{1}{2}} ;\,\,\, R=4$.
\item $c(n) =(2n)!/{(n+1)!}; \, \, \, \, {\tilde{W}}_5(x)  = -\frac{1}{2} + \frac{1}{\sqrt{\pi x}} e^{-\frac{x}{4}}+\frac{1}{2} {\rm erf} (\frac{\sqrt{x}}{2});\,\,\, R=\infty$; \\erf$(y)$ is the error function.
\item $c(n) ={(2n)!}/(n+1); \, \, \, \, {\tilde{W}}_6(x)  = \frac{1}{\sqrt{x}} e^{-\sqrt{x}} + {\rm Ei} (-\sqrt{x});\,\,\, R=\infty$;\\ Ei($y$) is the exponential integral.
\item $c(n) ={(3n)!}/{n!}; \, \, \, \, {\tilde{W}}_7(x)  = \frac{1}{3 \pi \sqrt{x}} K_{\frac{1}{3}}(2\sqrt{\frac{x}{27}});\,\,\,R=\infty$;\\ $K_{\nu}(y)$ is the modified Bessel function of the second kind.
\item $c(n)={(3n)!}/{(2n)!}; \, \, \, \, {\tilde{W}}_8(x)=\frac{\sqrt{3}}{27 \pi}\exp(-\frac{2x}{27})
[K_{\frac{1}{3}}(\frac{2x}{27})+K_{\frac{2}{3}}(\frac{2x}{27})];\,\,\,R=\infty$.
\item $c(n)={(3n)!}/{(n!)^3}$; (so-called middle trinomial coefficients);\, \, \, \, \\ ${\tilde{W}}_9(x)=\alpha {x}^{-\frac{2}{3}} {_2}F_{1}(\frac{1}{3},\frac{1}{3};\frac{2}{3};\frac{x}{27})+\beta {x}^{-\frac{1}{3}} {_2}F_{1}(\frac{2}{3},\frac{2}{3};\frac{4}{3};\frac{x}{27})\,$; \\where $ \alpha=\frac{1}{3}[\Gamma(\frac{2}{3})]^{-3}$ and $ \beta=-\frac{\sqrt{3}}{8\pi^{3}}[\Gamma(\frac{2}{3})]^{3}, \,\,\,R=27$.
\item $c(n)=\left(\begin{array}{c}3n\\n\end{array}\right)/(2n+1)$; \, \, \, \,  \\${\tilde{W}}_{10}(x)=\frac{\sqrt{3}\, 2^{\frac{2}{3}}}{12 \pi} 
\frac
{[2^{\frac{1}{3}}(27+3 \sqrt{81-12x})^{\frac{2}{3}}-6x^{\frac{1}{3}}]}
{x^\frac{2}{3}
(27+3 \sqrt{81-12x})^{\frac{1}{3}}}$;\,\,\, $R=\frac{27}{4}$.
\end{enumerate}
The above weight functions, which may be seen to be positive by inspection, constitute only a small selection of examples of this type.
Moreover, with the $c(n)$ of Examples 1--10, the normalization 
${\cal N}{_c}(x)$ of Eq.(\ref{norm}) can always be expressed in terms of known hypergeometric functions.  This renders calculations of many expectation values in the state $|z\rangle_c$ entirely analytic.

We now extend our considerations to include the important sequence of Bell numbers, $c(n)=B(n)$, which count the number of partitions of a set of $n$ objects.  The Bell numbers are given by the celebrated Dobinski formula\cite{com}
\begin{equation}
B(n) = 1/e \sum_{k=1}^{\infty} \frac{k^n}{k!}, \, \, \, \, n=0,1,\ldots,
\label{dob} \end{equation}
from which it readily follows that the solution of
$$\int_0^{\infty} x^n {\tilde{W}}_B(x) dx = B(n)$$ is 
\begin{equation}
{\tilde{W}}_B(x)=1/e \sum_{k=1}^{\infty}\delta(x-k)/{k!},
\label{wb}
\end{equation}
which is a sum of weighted Dirac delta function spikes situated on the positive integers.  Since $x=|z|^2$, the weight is positive on concentric equidistant circles around the origin of the complex plane. The particular form of Eq.(\ref{wb}) permits one to write down the weight function whose moments are $c(n) B(n),$
where $c(n)$ is any moment from Examples 1--10. For instance, if $c(n)$ are the Catalan numbers, the $n$th moment of   
\begin{equation}
{\tilde{W}}_{CB}(x)=1/(2\pi e) \sum_{k=1}^{\infty} {\frac{1}{k k!}}\sqrt{\frac{4k-x}{x}} H(4-\frac{x}{k})
\label{wbc}
\end{equation}
equals $(\left(\begin{array}{c}2n\\n\end{array}\right)/(n+1))\;B(n)$; $H(y)$ is the Heaviside function. The powerful convolution property of the Mellin transform\cite{six1,kla1,que} enables one to obtain Eq.(\ref{wbc}). It has been shown elsewhere\cite{gaz,hfu}  that by a different form of parametrization (replacing $z$ by two independent real variables) the states described above may be applied to specific physical systems. A Hamiltonian with energy levels $\varepsilon_n$ can be associated with the sequence $\{c(n)\}$ as follows:
\begin{equation}
\varepsilon_0 =0, \, \, \, \, \, \, \varepsilon_n=c(n)/c(n-1),\, \, \, n>0.
\end{equation}
The wealth of the preceding examples allows for different forms of $\varepsilon_n$, differing from the standard $\varepsilon_n =n$, associated with the standard coherent state characterized by the $|z\rangle_0$ of Eq.(\ref{cs}).
These examples are associated with models having discrete non-linear spectra, and may well provide good approximations to some real nonlinear systems.

\end{document}